\documentclass[sigconf,authorversion=true]{acmart}
\usepackage{booktabs}
\usepackage{url}
\usepackage{hyperref}
\makeatletter 
\g@addto@macro{\UrlBreaks}{\UrlOrds}
\makeatother

\setcopyright{acmcopyright}
\acmDOI{http://dx.doi.org/10.1145/3018896.3018918}
\acmISBN{978-1-4503-4774-7/17/03}

\acmConference[ICC '17]{Second International conference on Internet of Things, Data and Cloud Computing}{March 22--23, 2017}{Cambridge, United Kingdom} 
\acmYear{2017}
\copyrightyear{2017}

\acmPrice{15.00}

\begin{document}

\title{
  My Home is My Post-Office: Evaluation of a decentralized email architecture \\
  on Internet-of-Things low-end device
}

\author{Gregory Tsipenyuk}
\email{gregory.tsipenyuk@cl.cam.ac.uk}
\affiliation{%
  \institution{University of Cambridge Computer Laboratory}
  \streetaddress{William Gates Building, 15 JJ Thompson Avenue}
  \city{Cambridge}
  \country{United Kingdom}
  \postcode{CB3 0FD}
}
\author{Jon Crowcroft}
\email{jon.crowcroft@cl.cam.ac.uk}
\affiliation{%
  \institution{University of Cambridge Computer Laboratory}
  \streetaddress{William Gates Building, 15 JJ Thompson Avenue}
  \city{Cambridge}
  \postcode{CB3 0FD}
  \country{United Kingdom}
}

\begin{abstract}
Users predominantly access their email via mobile devices. This presents a two-fold challenge to the email applications. First, email's update from multiple devices has to be eventually reconciled with the server. Prioritization of updates is difficult and maybe undesirable. Solving this problem requires a data store with the complete history of email changes. Second, legacy email protocols don't provide an optimal email synchronization and access in mobile environment. In this paper we are proposing to take advantage of the Internet of Things (IoT) phenomena. In IoT environment a user may have multiple interconnected in-home low-end devices with publicly accessible address. In this architecture we move the email application from the central service into user's in-home and mobile devices, store complete email history on each device, and replace legacy IMAP and SMTP protocols with a synchronization protocol found in Distributed Version Control Systems(DVCS). This addresses the email reconciliation issue, optimizes the bandwidth usage, and intrinsically puts the user in control of her data. We analyze a number of stores and synchronization implementations and compare them with the open source Dovecot email server.
\end{abstract}


\keywords{Internet-of-things, eventual consistency, revision control, decentralization, email}

\maketitle

\section{Introduction}
\label{s:introduction}

Email is ranked as one of the top internet activities~\cite{BrandonGaille2013}~\cite{McGee2013}. It is presently provided by centralized services running in the cloud. The centralized architecture is required for the following reasons:
\begin{itemize}
\item
 To access email by clients located behind middle-boxes with no globally accessible address. The central service allows a client to connect to the server via the domain name, which is resolved by Dynamic Name System (DNS) to public IP address.
\item
 The sender and the recipient usually are not on-line at the same time. The store-and-forward property enables the sender to send the email to the central server where it is later retrieved by the recipient.
\item
 The central service provides high availability via redundancy and replication.
\item
 The central service backs up email archives.
\end{itemize}
While the email is free, there are associated intangible costs to users in terms of privacy, where the email provider data-mines user's email, and the user's account vulnerability to hacking~\cite{Kopstein2013}~\cite{Madden2014}.

According to Edwards W.K., et.al. in Bayou, electronic mail is often considered to be the 'classical' asynchronous collaborative application~\cite{Edwards1997a}. This type of network shared-data system according to Brewer's CAP theorem is characterized by high availability and tolerance to network partitions with eventually consistent database~\cite{Brewer2000}. These properties are inherent in delay tolerant email's store-and-forward architecture. Generally, an email client, maintains the cache of messages. The cache is synchronized with the server by legacy IMAP~\cite{Crispin2003} and to some extent SMTP~\cite{Klensin2001} protocols. The client uses IMAP protocol to validate the cache via combination of mailbox's statistics and unique keys. Changes made to the same mailbox by multiple clients can result in communication overhead to identify divergence point and may still cause the complete mailbox synchronization. A number of protocol extensions, for instance MODSEQ~\cite{Melnikov2008} and IDLE~\cite{Leiba1997},  have been defined to improve the communication efficiency, especially in resource-limited mobile devices.

Schmandt and Marti pointed out in 2005 that mobile email usage is growing fast with increasingly heterogeneous multi-device access to the email~\cite{Schmandt2005}. In fact, the mobile email access overtook the desktop in 2011\footnote{https://www.campaignmonitor.com/dev-resources/will-it-work/email-clients/}. This transition to mobile computing along with the email overload may have affected user's email behaviour. In ~\cite{Castro2016} it is shown that 89.5\% of the email delete actions are delete-without-read. I.e. users delete the email without even opening, let alone reading it. Authors suggest that this phenomena could be explained by increased number of machine-generated email, which accounts for 90\% of non-spam Web email~\cite{Grbovic2014}. It is also possible that users are checking their email on mobile devices while busy with other activities and have less patience to read the entire message, deleting it based on cues such as the message's subject or preview.

We therefore see several issues due to the email access from multiple intermittently connected devices. First, multiple email changes are reconciled, with the most recent update overwriting the previous ones. This maybe undesirable or simply not what a user wants. Second, a user may unintentionally delete an important message or file the message to the wrong or obscure folder. Finally, the legacy IMAP protocol is not bandwidth efficient. To address these issues it is not sufficient to have the latest email state. As Brewer notes ``The state is less useful than the history, from which the system can deduce which operations actually violated invariants and what results were externalized, including the responses sent to the user'' and ``The best way to track the history of operations on both sides is to use version vectors, which capture the causal dependencies among operations''~\cite{Brewer2012}. The replica of the history can be maintained with DVCS like synchronization protocol designed for the bandwidth optimization via efficient divergence point search and content de-duplication.

Recent phenomena of the Internet of Things (IoT) will see the number of interconnected devices grow to 24 billion by 2020~\cite{Gubbi2013}. A device could be a home router or an electricity monitor. While a resource limited, some of these devices, like Raspberry Pi\footnote{https://www.raspberrypi.org}, are comparable in the hardware configuration to an average smartphone. We see the IoT environment as an opportunity to decentralize the email. 

The contribution of this paper is four-fold. First, we are proposing a high level decentralized email architecture where the email history is stored on a cluster of user's owned devices. The history addresses possible inconsistencies and user's error due to the email access from multiple devices. Moving the data to user's devices intrinsically solves the privacy issue. The cluster provides availability, redundancy, and backup. Email replicas on devices are maintained via efficient DVCS like synchronization
protocol. Second, we present a detailed evaluation of the email architecture on Raspberry Pi computer. Third, the evaluation shows that the approach is both feasible and affordable. Fourth, distributed architecture evolves out of 1) taking a modern view of what an email architecture requirements are, including eventual consistency for synchronization, and modern approaches (i.e. CAP) to consistency; 2) the advent of IoT, solving the reachability and ubiquity problem of availability of
peer-to-peer (P2P).

The rest of the paper is organized as follows. Section 2 reviews the related work. Section 3 provides high level architecture of the proposed email system. Section 4 describes the evaluation design. Section 5 presents evaluation data. Finally, section 6 concludes.

\section{Related Work}
\label{s:relatedwork}

Earlier email research has been focusing on achieving high availability and reliability via replication of the delivery path as in Grapevine~\cite{Birrell1982}, distributed file system as in Andrew Message System~\cite{Rosenberg1987} and Porcupine~\cite{Saito1998}, and clustering like in NinjaEmail~\cite{VonBehren2000} and Porcupine. 

The Bayou system~\cite{Edwards1997a} has a different approach with an emphasis on a replicated weakly-consistent storage system. Bayou also introduces ``Timewarp'', a toolkit, which provides the versioning functionality. 

The turning point in the email research correlates with introduction and success of P2P file sharing systems like Napster\footnote{https://en.wikipedia.org/wiki/Napster}, Gnutella\footnote{https://en.wikipedia.org/wiki/Gnutella}, and BitTorrent\footnote{https://en.wikipedia.org/wiki/BitTorrent}. In P2P, participating peers contribute their resources to solve some common task. P2P is characterized by high degree of decentralization, self-organization, organic growth, resilience to faults and attacks, and abundance and diversity of resources~\cite{Rodrigues1999}. 

While there are implementation nuances, generally, a P2P email architecture relies on Distributed Hash Table (DHT)\footnote{https://en.wikipedia.org/wiki/Distributed\_hash\_table} to replicate the content between peers. For instance in ~\cite{Kangasharju2003}, system nodes (super nodes) in DHT provide persistence of messages in transit from the sender to the receiver. The storage is not durable and messages are deleted after reading. 

ePost~\cite{Mislove2006a} has a durable DHT storage and in addition uses DHT as the multicast notification system. ePost maintains user's view, i.e. mailboxes, as the history of changes in an immutable log with periodic snapshots for fast traversal. 

~\cite{Kageyama2008} is a spam-avert, pull-based system with the initial storage burden placed on the sender's trusted group of peers and receiver using DHT notifications to retrieve the email or delete it without downloading.  

DMS~\cite{Emanuel2011} uses peer's resource self-evaluation for classification into hierarchy of nodes, each maintaining and replicating the data according to its function. 

Similarly,  HMail~\cite{Mezo2012} builds a hierarchy of overlays, with one group of peers having higher uptime and bandwidth, and another group having higher processing power and storage. HMail also takes into consideration peer's geographical location. 

~\cite{Zhao2004} is a hybrid P2P where depending on the environment and capabilities super nodes have functionality of the centralized server, delegate tasks to regular nodes, or maintain inbox and outbox with references to message's location. The replication hash algorithm evaluates peer's on-line habits, workload, and trust relationship to select optimal nodes. 

In Decentralized Electronic Mail (DEM)~\cite{Bercovici2006} user's mailbox is a mobile object replicated over participating nodes via a middleware rather than DHT. The attachment is the mobile object too, consequently only one, albeit replicated, unique copy of the attachment is maintained. Mail items travel directly from the sender to the receiver achieving O(1) communication cost. 

Apache Wave (originally Google Wave)\footnote{https://en.wikipedia.org/wiki/Google\_Wave\_Federation\_Protocol} is an ambitious architecture merging instant messaging, email, wikis, and social network under the web-based computing platform. Apache Wave messages (waves) along with their history are perpetually stored on a central server. Waves support concurrent modification and low-latency update and are shared with collaborators. Google Wave was not successful, with the failure attributed to overly complicated interface yet without any apparent benefit over existing solutions\footnote{http://arstechnica.com/information-technology/2010/08/google-wave-why-we-didnt-use-it/}.

While P2P architecture has attractive properties outlined above, it requires a complex adaptive infrastructure. There are successful P2P applications like BitTorrent or Bitcoin\footnote{http://s.kwma.kr/pdf/Bitcoin/bitcoin.pdf}, but to-date P2P found no traction in the email outside of academia interest.

\section{High level email architecture}
\label{s:arcitecture}

\begin{figure}
  \centering
  \includegraphics[width=\columnwidth]{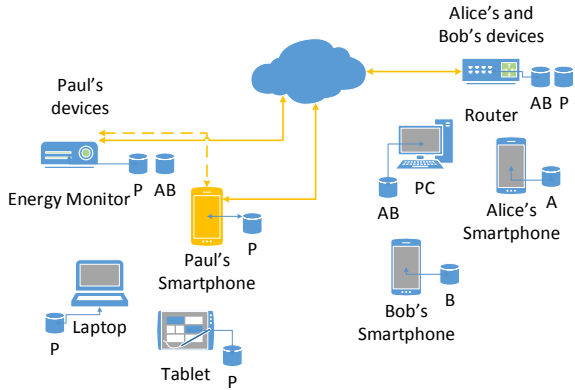}
  \caption{\label{f:architecture}High level email architecture on in-home IoT devices.}
\end{figure}

Our architecture is based on the previous research where the email history persistence is one of the features, most notably Google Wave and to some extent Bayou and ePost, and the research into the email decentralization via P2P network. The main inspiration for our research is owned to IoT phenomena where a user may have at her disposal multiple interconnected devices. We are particularly interested in devices that generally are powered on for extended period of time and plugged into AC power supply, for instance a Network Router or  Electricity Monitor. User's devices create a trusted group of peers with shareable resources to provide availability, redundancy, back up, and intrinsically privacy. We assume that at least one device has a globally accessible address and other peers can connect to this device. 

Figure~\ref{f:architecture} shows the high level architecture's example. We consider two peer groups. First group consists of Paul's devices and the second group of Alice's and Bob's devices. Each group has at least one globally accessible device, Energy Monitor in the first group, and Router in the second group. Each device maintains complete history of the email and synchronizes its replica with the master replica on the globally accessible device. Each personal device only maintains the
replica of its owner, for instance Alice's smartphone only keeps Alice's email. Globally accessible devices and other devices shared within the group maintain email of all group members. For instance, Router and PC maintain both Alice's and Bob's email. Alice and Bob are socially connected to Paul and actively exchange emails with each other. Their Globally accessible devices maintain replica of both of their email archives providing higher availability, redundancy, backup, and efficient disk
usage. For instance, Paul can access his email via his smartphone from either one of the globally accessible devices (solid yellow connector). When at home, Paul can access his email by directly connecting to his device via WiFi (dashed yellow connector). Sharing resources between socially connected group of peers like family members or close friends reduces the network and disk energy cost due to the attachment's de-duplication. We analyzed attachment's statistics in the Enron email corpus
data\footnote{https://www.nuix.com/edrm-enron-data-set/enron} consisting of 130 user accounts with the total size of unique messages equal to 19.9GB. Attachments in average contribute 67.79\%, with duplicate attachments contributing 12.16\% to each user's account. Duplicate attachments between all user accounts take up 18.19\% of the space. We also analyzed 20 email accounts from friends and family totaling 50.48GB. The statistics is on the same order of magnitude: 75.13\%, 10.96\%, 25.98\%. If
all our family members keep their email on the same devices then there is about 25\% saving on the disk space, network bandwidth, network, and disk IO energy between all members. As we'll show later in the paper, there is a linear dependency between the network and disk IO energy and the size of the archive. Clearly, there are some family members that share more attachments than the others. Social network analysis can be used to discover the best match between the family members to optimize the cost of device sharing. This is a subject for future research.

\section{Evaluation design}

We run our evaluation on the latest release 3 of Raspberry Pi. Raspberry Pi is a credit card-size single board computer developed in the UK with the intent to promote the teaching of basic computer science. We have chosen Raspberry Pi for the following reasons, which make it an attractive platform for developing IoT applications to both a community of enthusiasts and large companies like Microsoft and IBM\footnote{https://www.raspberrypi.org/blog/tag/internet-of-things/,http://www.informationweek.com/software/enterprise-applications/10-raspberry-pi-projects-for-learning-iot/d/d-id/1320757, https://developer.microsoft.com/en-us/windows/iot, http://www.ibm.com/internet-of-things/ecosystem/devices/raspberry-pi/}.
\begin{itemize}
\item
Small form factor
\item
Capable hardware platform with 64-bit quad-core ARMv8 1.2GHz CPU, 1 GB RAM, micro SD-card up to 128 GB, 802.11n Wireless LAN, Bluetooth, Ethernet port, 4 USB ports, Full HDMI port, 40 GPIO pins, Micro USB power supply
\item
Powerful Linux development environment with officially supported Raspbian Jessie OS, based on Debian Jessie OS
\item
Large support community with over five million devices sold since its first release in 2012
\item
Affordable \$35 price tag  
\end{itemize}

Authors in ~\cite{Hameed2015} evaluate Raspberry Pi as an affordable, lightweight, and energy-efficient private email infrastructure. Authors conclude that Raspberry Pi is an adequate platform for individual or small and medium enterprises with up to 4000 email load per day. This analysis encouraged us to go on with the evaluation of Raspberry Pi as the hardware platform in the email architecture supporting revisions, data replication and efficient synchronization.

\begin{figure}[h]
  \centering \includegraphics[width=\columnwidth]{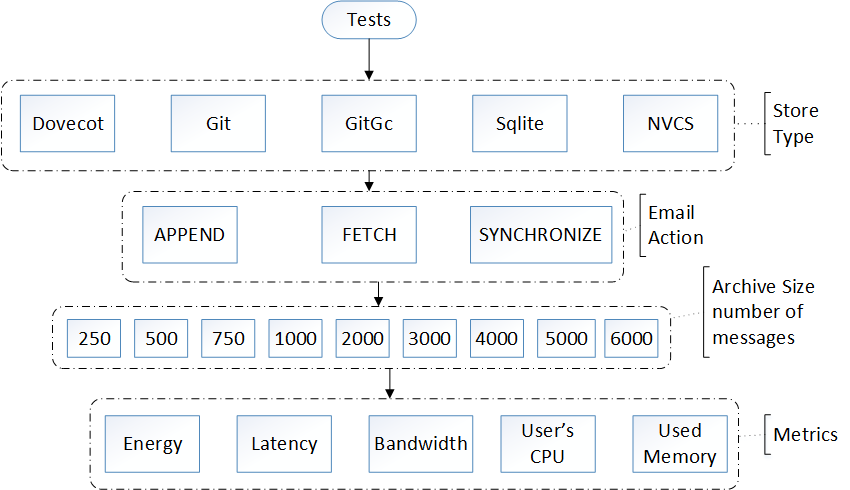}
  \caption{\label{f:metrics}Evaluation workflow.}
\end{figure}

Figure~\ref{f:metrics} shows evaluation's workflow. We evaluate five email back-end store types. First is Dovecot version 2.2.24\footnote{http://www.dovecot.org}, open source high performing IMAP server, which we use as the base system in our evaluation. We use Maildir\footnote{https://cr.yp.to/proto/maildir.html} as the email data structure. Second is Git\footnote{https://git-scm.com}, a DVCS system with emphasis on performance, non-linear workflow, and efficient synchronization. Revision maintenance and synchronization in Git is done via
Merkle\footnote{https://en.wikipedia.org/wiki/Merkle\_tree} hash tree. Third is GitGc, this is Git with garbage collection run after each appended 250th message. Fourth is Sqlite library database, which is the most widely deployed database, supporting iOS, Android, and embedded applications. We maintain revisions in Sqlite via audit tables. Sqlite maintains the database in a single file. Finally, we have a naive VCS implementation (NVCS), with revisions maintained via single log file constructed as Merkle tree blockchain. We used logical Maildir structure in Git, GitGc, Sqlite, and NVCS. To model store types, we implemented in C++ a TCP/IP server with support for IMAP's APPEND and FETCH commands. We compressed messages on disk and used fdatasync to force all modified in-core data to be written to disk.

For each store type our evaluation assesses three email actions: append, fetch, and synchronize. Each action is tested five times for an archive with 250, 500, 750, 1000, 2000, 3000, 4000, 5000, and 6000 number of messages and archive's size 41.18, 54.18, 64.85, 122.65, 256.4, 442.8, 574.27, 604.05, and 831.74 MB respectively. Each archive is randomly generated from the Enron corpus email dataset. We use Mac OS X MacBook Air 1.7GHz, 8GB RAM, and 512 SSD to append and fetch email messages and as
the remote repository for synchronization with the local Raspberry Pi repository. Synchronization is tested by first appending unique 100 messages to already created archives and then synching the archives to the original archives. Raspberry Pi is connected to Mac via Netgear 10.100M FS608 switch. For each test we collect energy, latency, bandwidth, user's cpu, and used memory. The energy and latency are tracked with Monsoon Power Monitor FTA22D. CPU and memory are tracked with
top\footnote{http://man7.org/linux/man-pages/man1/top.1.html} Linux utility. We pick the overall maximum user's CPU. To get the memory we calculate the difference between the lowest and the highest memory used during the execution, where the memory used is calculated as (mem-free-buffers-cached). The bandwidth is calculated from tcpdump output by adding up payload packet's length. Error bars on each plot show 95\% confidence intervals, which are illegible in some cases because of low variance in
measurements. Large dots on all plots represent average values and small-dotted lines represent trend line, which are always moving averages for CPU and memory.

\section{Evaluation}

\subsection{Appending Messages}
\label{s:appending}

\begin{figure}[h]
  \centering \includegraphics[width=\columnwidth]{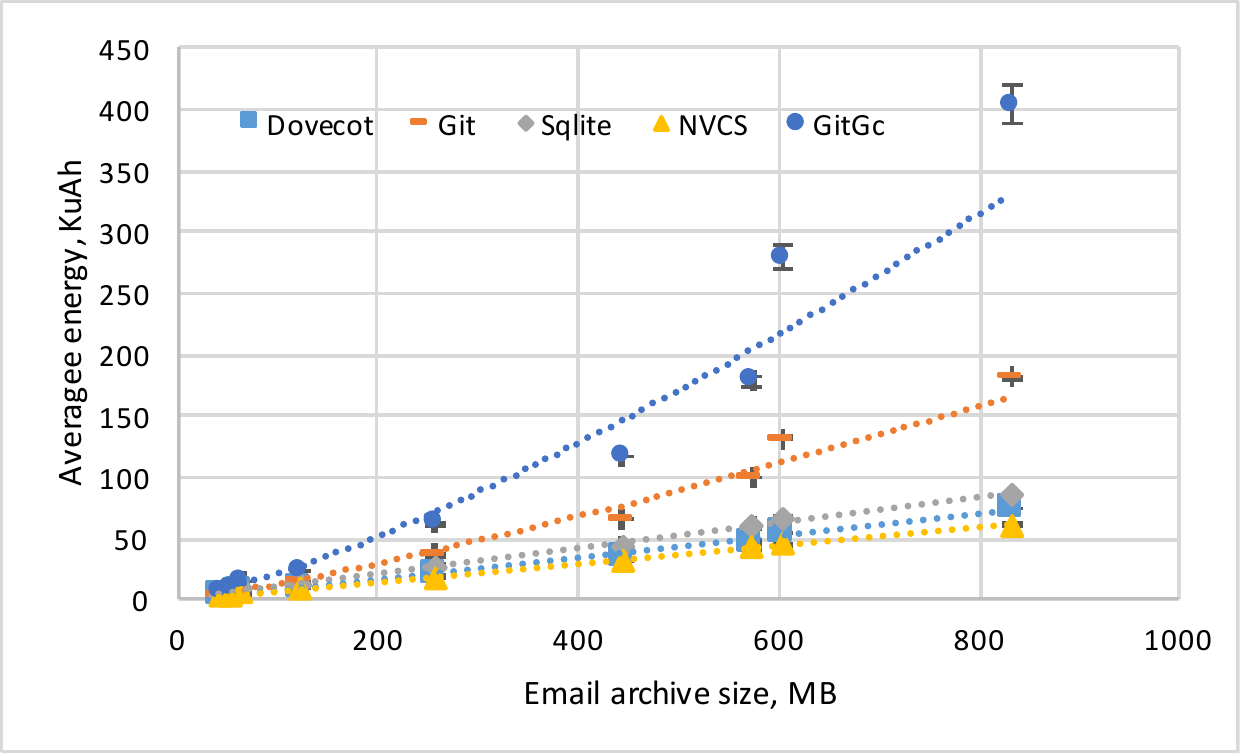}
  \caption{\label{f:appendenergy}Append messages, average energy.}
\end{figure}
\begin{figure}[h]
  \centering \includegraphics[width=\columnwidth]{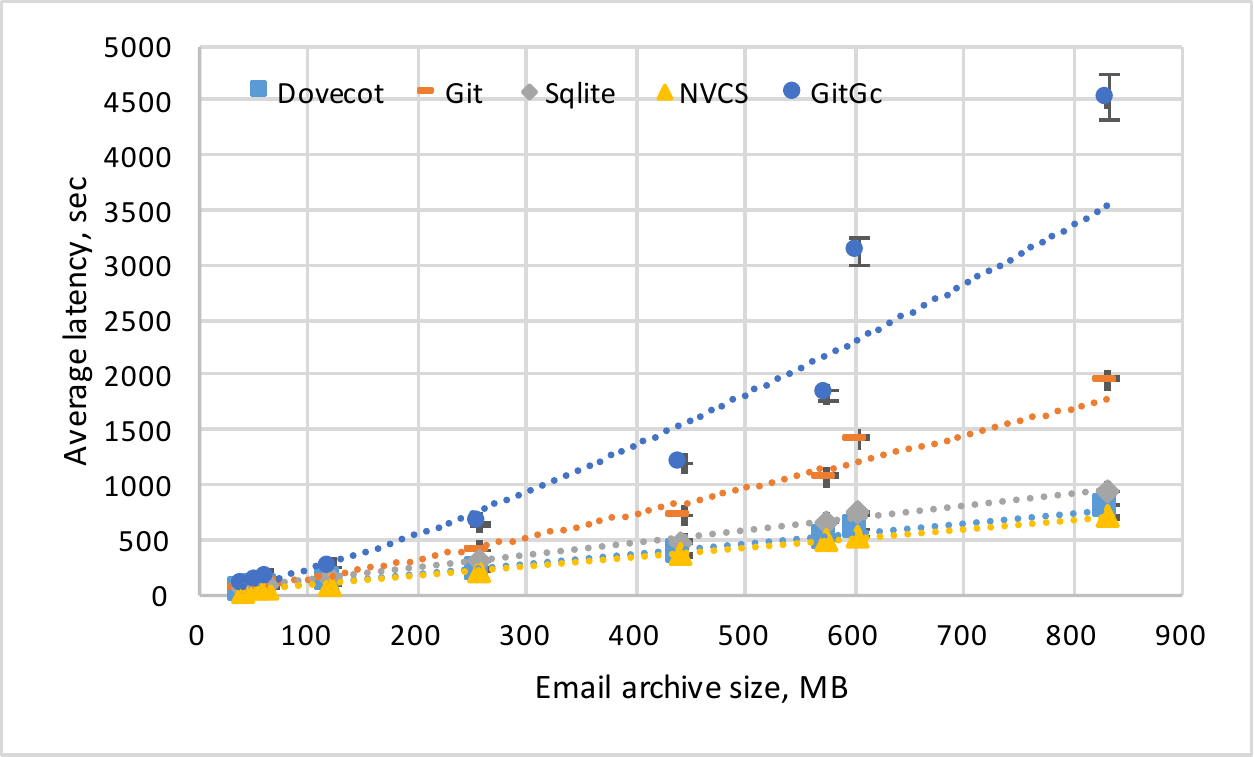}
  \caption{\label{f:appendlatency}Append messages, average latency.}
\end{figure}
\begin{figure}[h]
  \centering \includegraphics[width=\columnwidth]{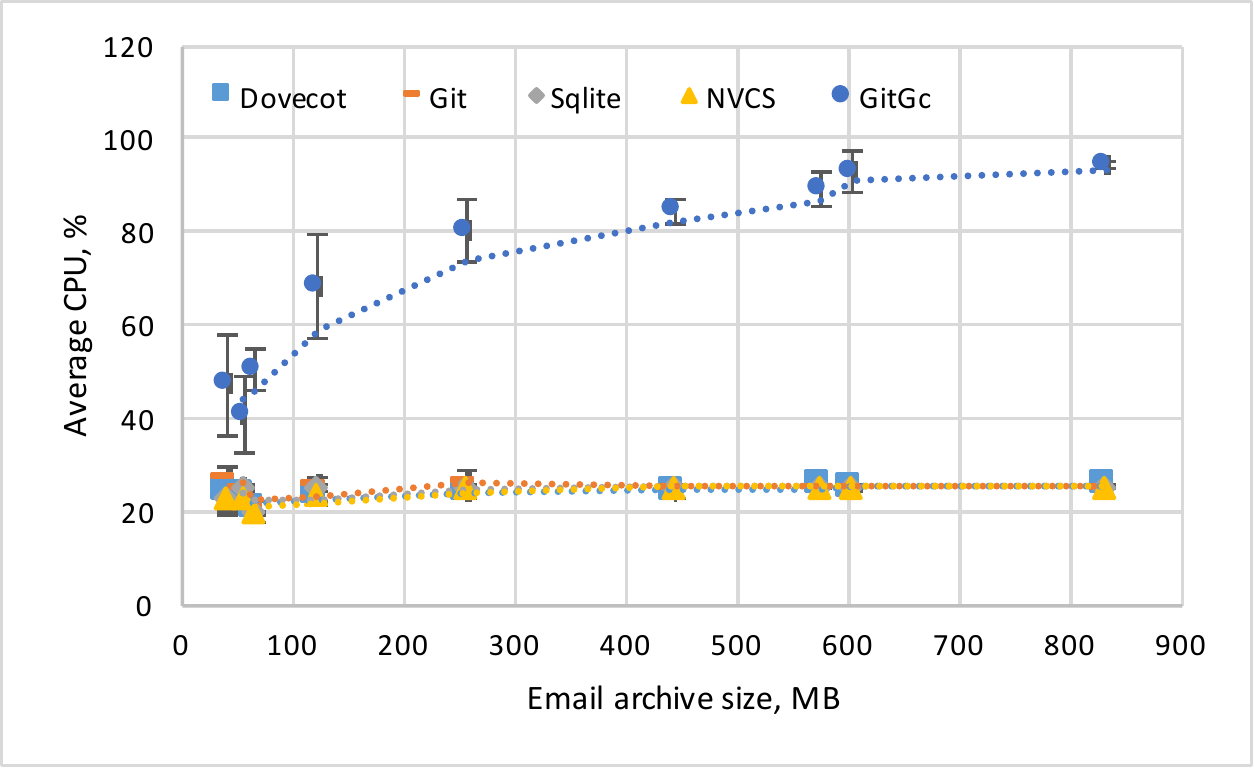}
  \caption{\label{f:appendcpu}Append messages, average user CPU.}
\end{figure}
\begin{figure}[h]
  \centering \includegraphics[width=\columnwidth]{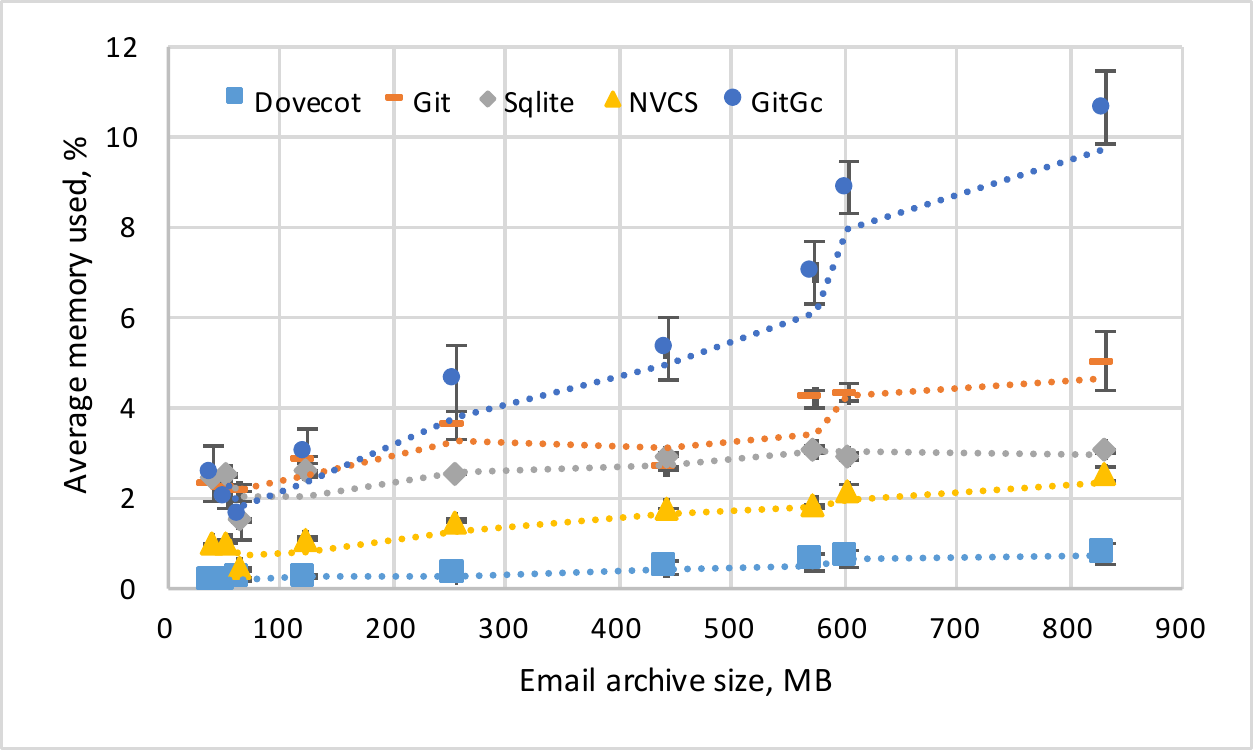}
  \caption{\label{f:appendmemory}Append messages, average user memory.}
\end{figure}

We first analyze results of appending a set of messages via IMAP APPEND command. Figures ~\ref{f:appendenergy},~\ref{f:appendlatency},~\ref{f:appendcpu},~\ref{f:appendmemory} show the energy, latency, CPU, and memory depending on the appended archive size. Energy and latency ranking best to worst is NVCS, Dovecot, Sqlite, Git, and GitGc. Intuitively we expected NVCS, Dovecot, and Sqlite perform similarly, and it is confirmed by the plots. Indeed, they all have in common three computational tasks: network read, compression, and disk write plus some overhead. Where the overhead varies between the stores and can be explained by NVCS's SHA1 computation and log file creation, Dovecot's user account management, and Sqlite database's management, with NVCS having the lowest overhead. 

Git and GitGc have the same tasks but significantly higher energy and latency overhead, which is explained by Git maintaining revisions via snapshots, with rough disk space cost O(\(N^2\)). Indeed, the energy and latency trend line for Git and GitGc is O(\(N^2\)) while other stores trend line is O(N). GitGc has even higher overhead than GitGc for two reasons. First, the garbage collection runs delta compression to pack loose files into one single file. This process is CPU and memory demanding, which is confirmed by CPU and memory usage on Figure ~\ref{f:appendcpu},~\ref{f:appendmemory}. Second, the cost of disk IO is the same as in GitGc plus the overhead of creating the pack file. GitGc compression is very efficient with the resulting pack file having fairly small disk overhead, but there is a significant temporary disk overhead because the new pack is written to the temporary file first and then deleted after the new pack is completed.
Unlikely from GitGc, other stores are not CPU bound, with CPU almost constant around 25\%. Memory usage has a slight upward trend but is well below 6\%. The bandwidth usage is the same for all stores. Indeed, the same batch of messages is uploaded to the servers regardless of the store type.

\subsection{Fetching Messages}
\label{s:fetching}

\begin{figure}[h]
  \centering \includegraphics[width=\columnwidth]{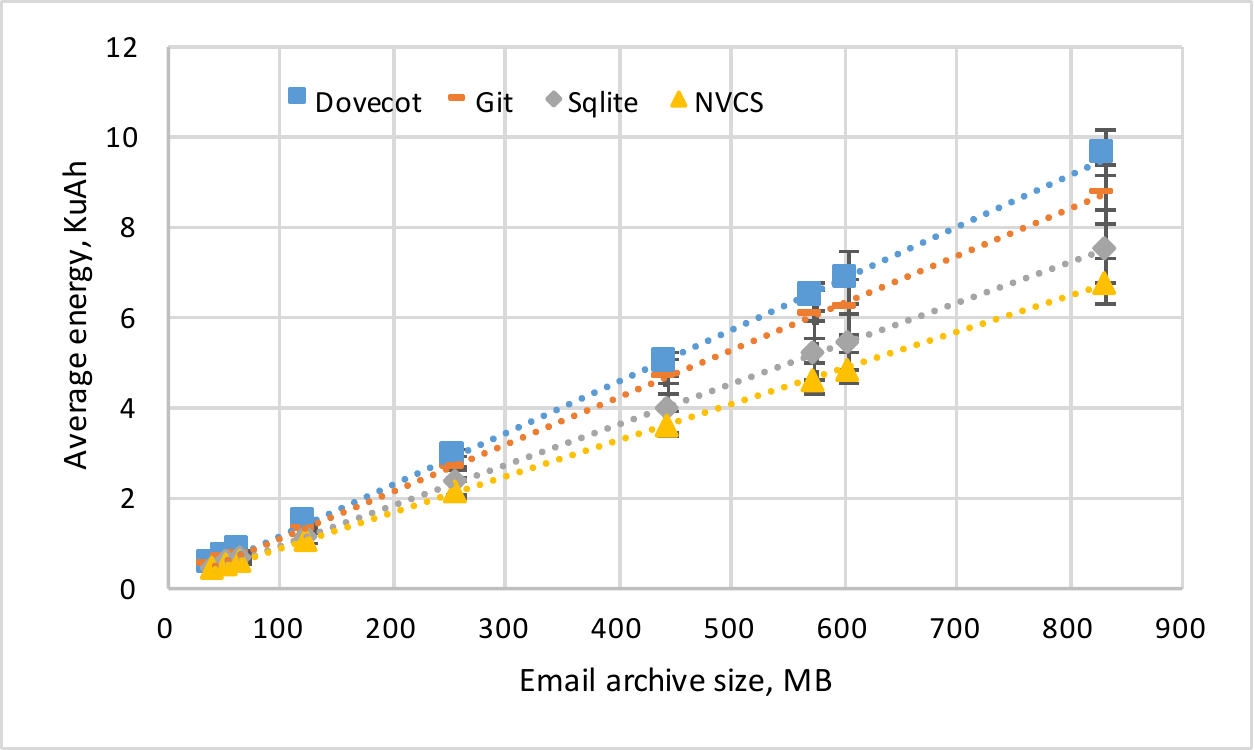}
  \caption{\label{f:fetchenergy}Fetch messages, average energy.}
\end{figure}
\begin{figure}[h]
  \centering \includegraphics[width=\columnwidth]{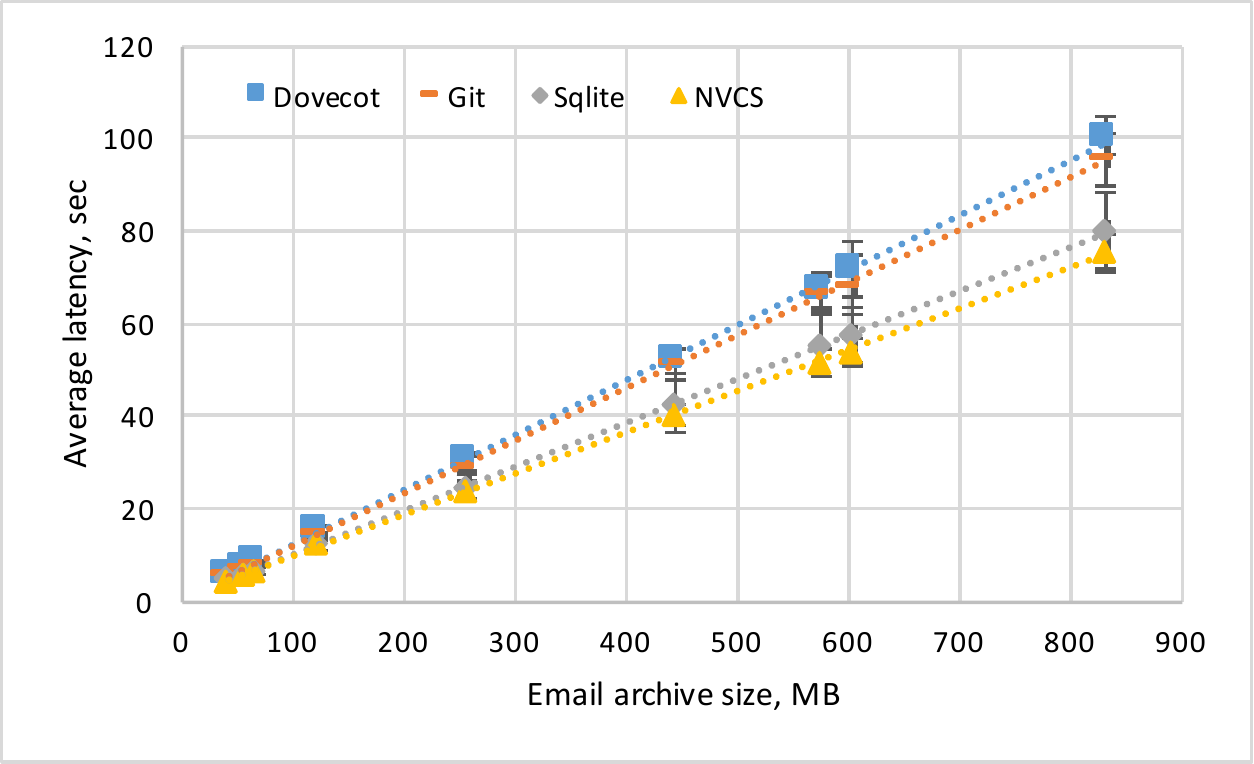}
  \caption{\label{f:fetchlatency}Fetch messages, average latency.}
\end{figure}
\begin{figure}[h]
  \centering \includegraphics[width=\columnwidth]{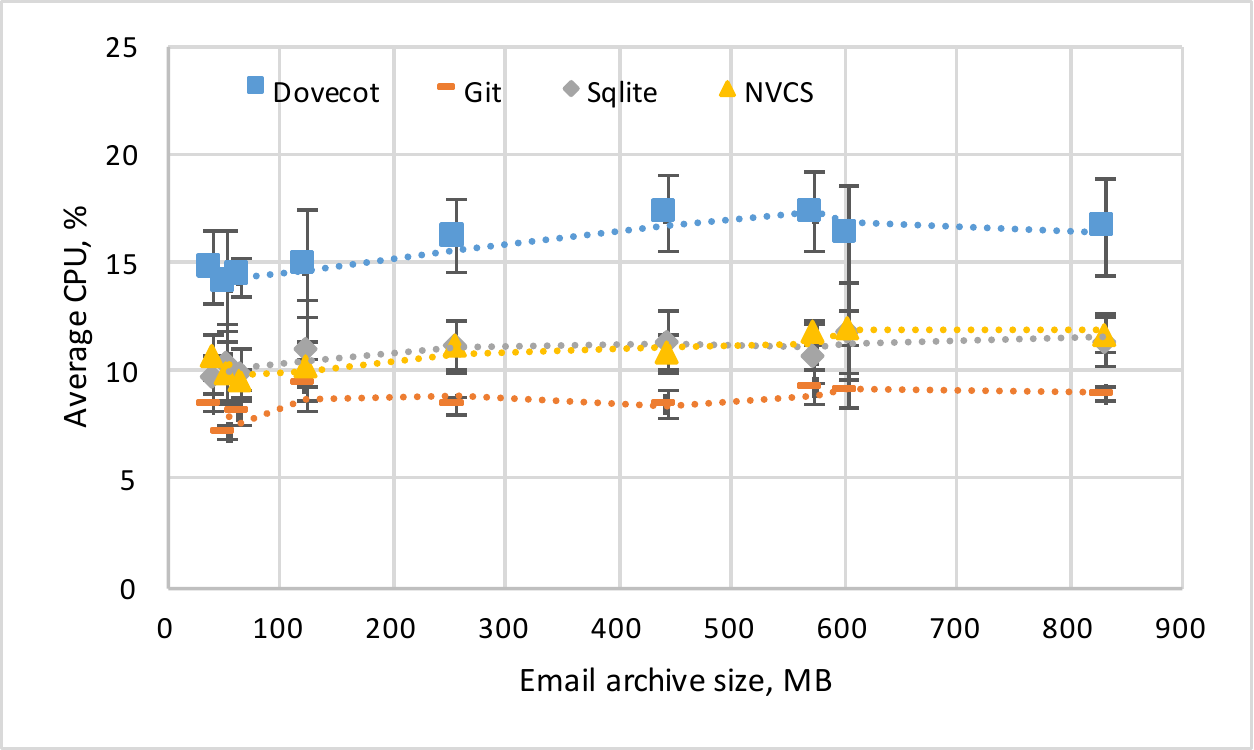}
  \caption{\label{f:fetchcpu}Fetch messages, average user CPU.}
\end{figure}
\begin{figure}[h]
  \centering \includegraphics[width=\columnwidth]{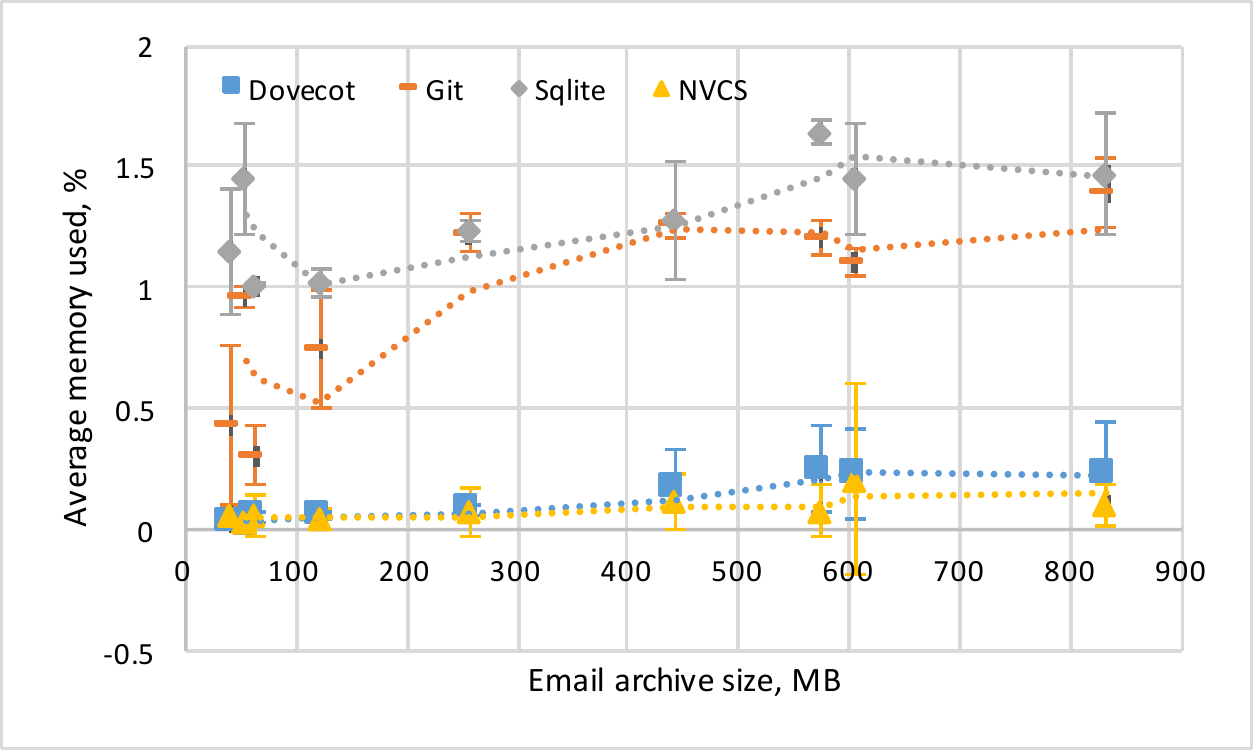}
  \caption{\label{f:fetchmemory}Fetch messages, average user memory.}
\end{figure}

Figures ~\ref{f:fetchenergy},~\ref{f:fetchlatency},~\ref{f:fetchcpu},~\ref{f:fetchmemory} show the energy, latency, CPU, and memory depending on the appended archive size. We used IMAP FETCH command to fetch all messages. We see less variation in energy and latency between the stores even for Git (we don't show GitGc plots because they are practically identical to Git). Indeed, all stores have essentially the same process to fetch the messages: 1) read index file; 2) get the message location; 3) read the message; 4) decompress the message; 5) write the message to the network. While there are some variations in CPU and memory, none of the stores is either CPU or memory bound with the CPU usage under 20\% and the memory usage under 2\%.
Bandwidth usage is the same in all cases since once the message is red from the disk and uncompressed, it produces the same size regardless of the store type.

\subsection{Archive Synchronization}
\label{s:synchronization}

\begin{figure}[h]
  \centering \includegraphics[width=\columnwidth]{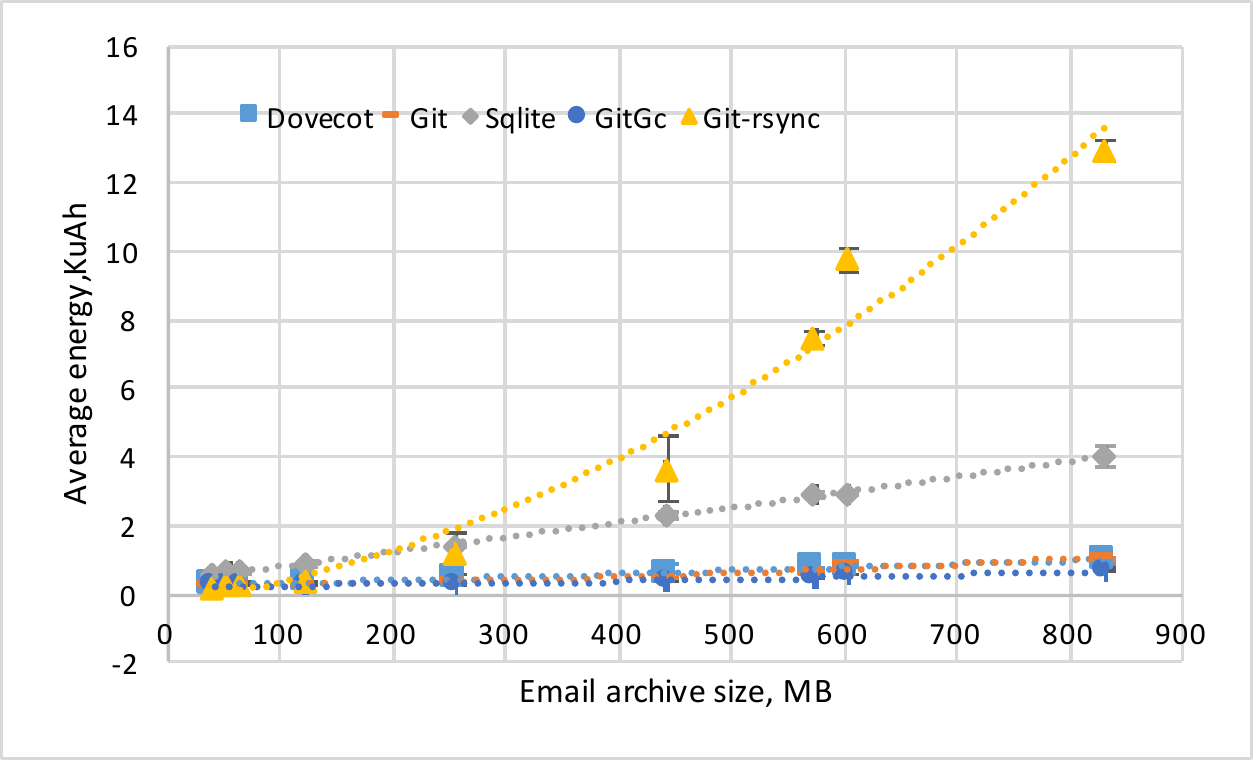}
  \caption{\label{f:syncenergy}Sync archives, average energy.}
\end{figure}
\begin{figure}[h]
  \centering \includegraphics[width=\columnwidth]{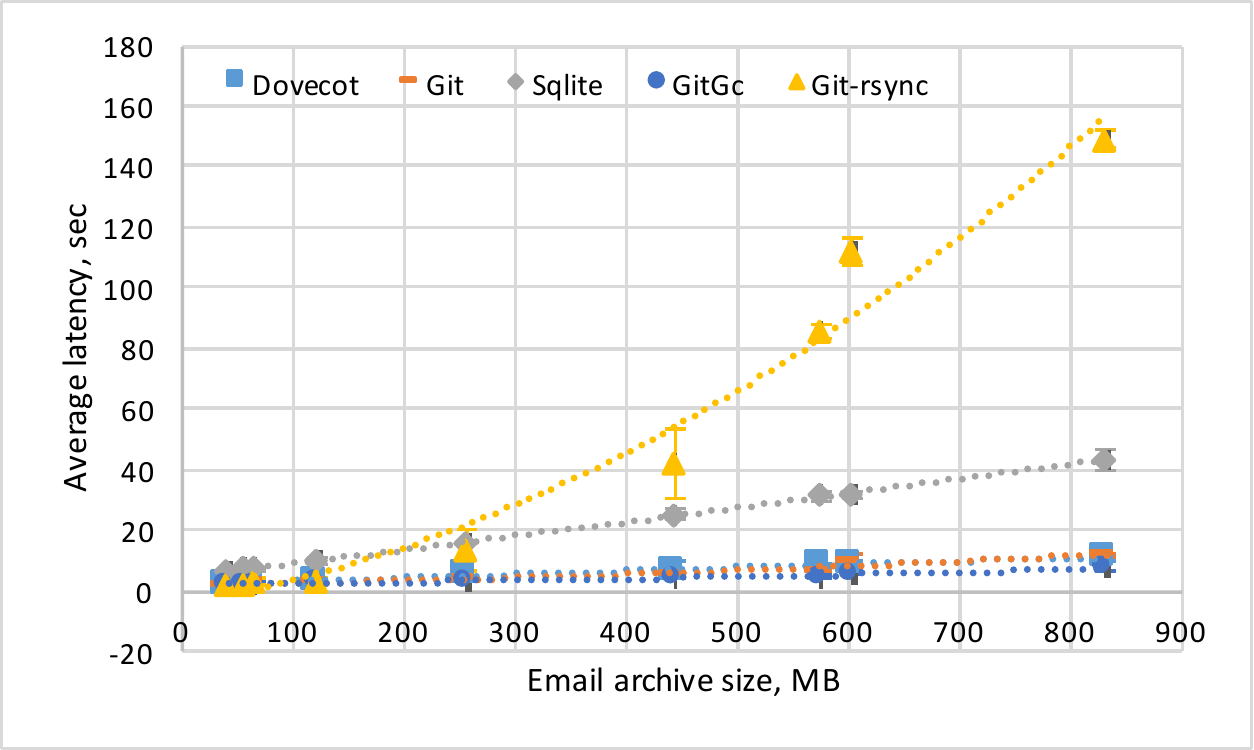}
  \caption{\label{f:synclatency}Sync archives, average latency.}
\end{figure}
\begin{figure}[h]
  \centering \includegraphics[width=\columnwidth]{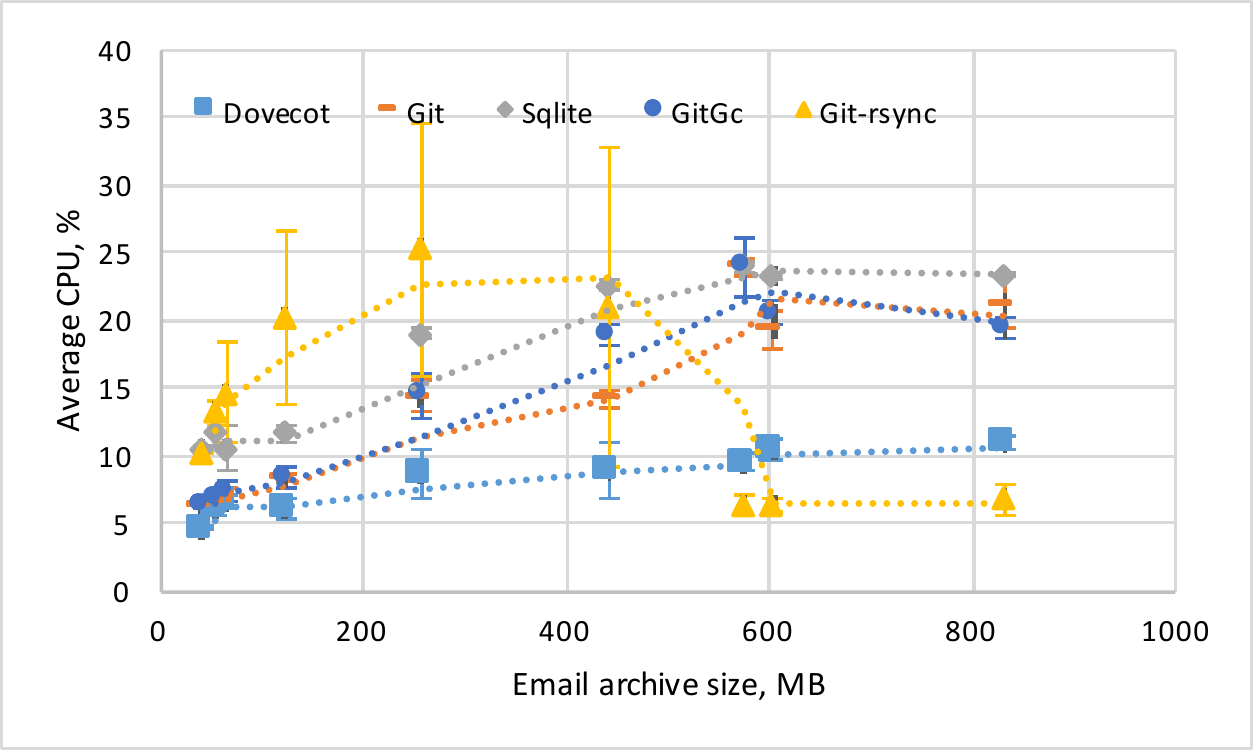}
  \caption{\label{f:synccpu}Sync archives, average user CPU.}
\end{figure}
\begin{figure}[h]
  \centering \includegraphics[width=\columnwidth]{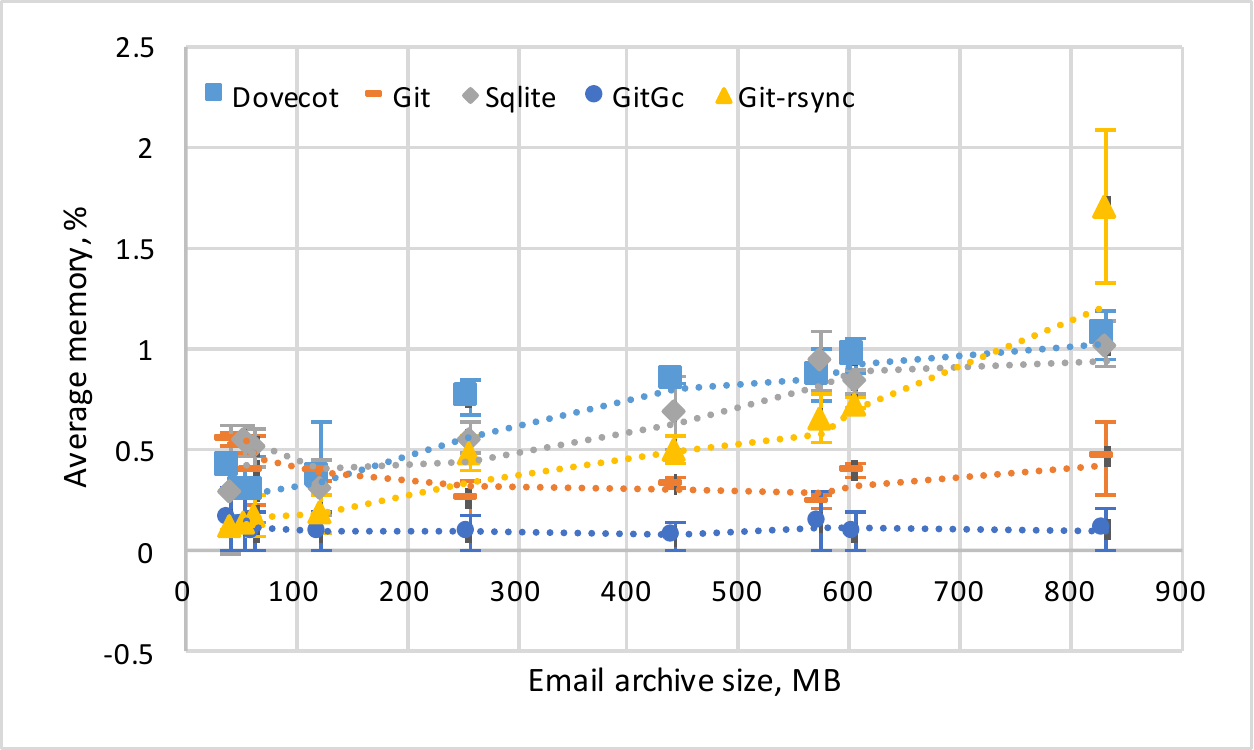}
  \caption{\label{f:syncmemory}Sync Archives, average user memory.}
\end{figure}

\begin{figure}[h]
  \centering \includegraphics[width=\columnwidth]{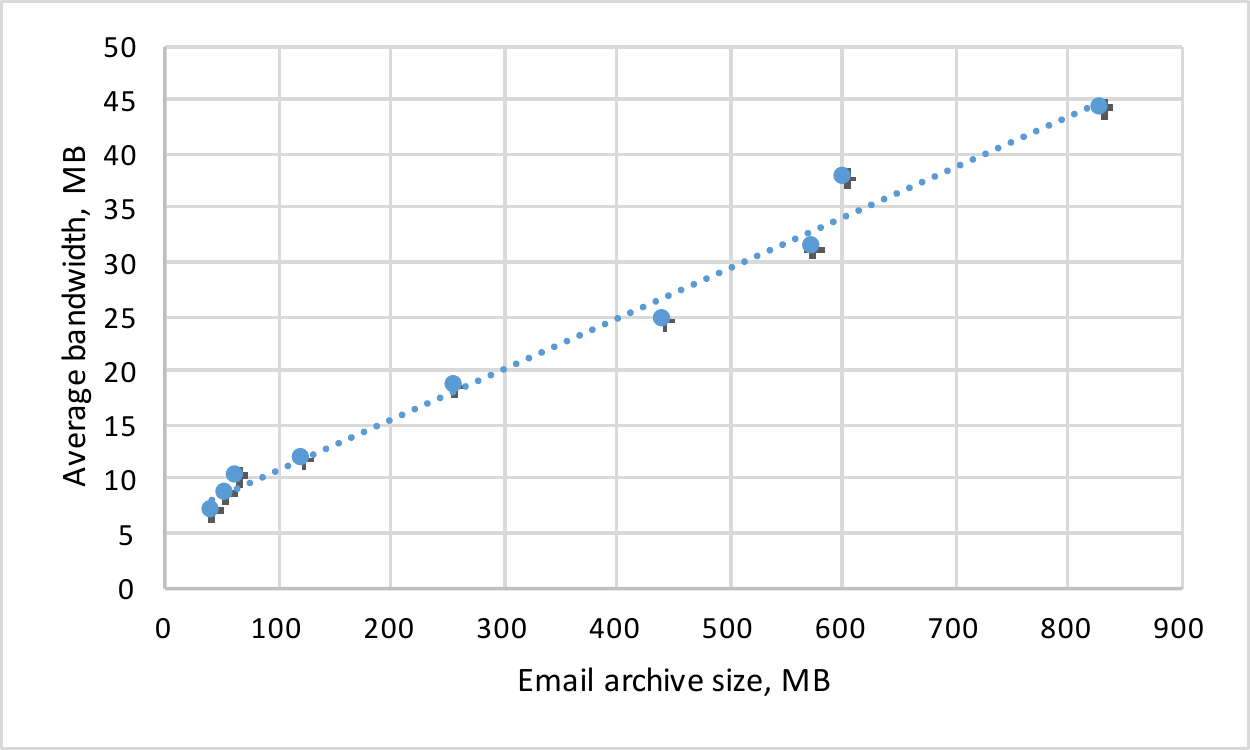}
  \caption{\label{f:bandwidth}Average bandwidth usage in Git-rsync.}
\end{figure}

Figures ~\ref{f:syncenergy},~\ref{f:synclatency},~\ref{f:synccpu},~\ref{f:syncmemory} show the energy, latency, CPU, and memory depending on the appended archive size. We used Dovecot's dsync utility, rsync utility for Sqlite and Git-rsync, and Git's(and GitGc) fetch to synchronize local to remote archives. Git-rsync highlights the difference between rsync for single file (Sqlite) versus multiple files (Git) synchronization and the difference between rsync and Git synchronization algorithm.
Rsync performs the worst in all measurements, with multiple files (Git-rsync) being worse than single file (Sqlite) synchronization. In case of Git-rsync, rsync has to scan multiple files (\(N^2\)) to find updated or new files to transfer. This results in high volume of the control information exchanged between the local and remote servers as can be seen from the bandwidth usage shown on Figure ~\ref{f:bandwidth}. In case of Sqlite, rsync scans for file chunks, which changed between the local
and remote file, and transfers only changed chunks. Rsync can guess location of the change based on the file size to reduce the volume of control information. Rsync creates a new copy of the file, moving it into place when the transfer is complete. This is a default behavior over the in-place update. The former is clearly preferred as in the latter case an update failure will leave the entire email archive unusable. We also used rsync to synchronize Dovecot archives, which have O(N) files. In this case rsync performed better than Dovecot's dsync utility but worse than GitGc. We don't show Dovecot rsync plots to reduce the figure's clutter.
GitGc performs the best. The reason for this is most likely due to the efficient scan of divergence point in the Merkle hash tree in the single indexed pack file rather than scanning multiple files. Git and Dovecot have practically the same performance.
The best trend line match for energy and latency in all stores is second order polynomial. It is a slightly better match over the linear trend line. For instance, \(R^2\) for GitGc is 0.981 for the polynomial versus 0.951 for the linear trend line. Since the network and disk write have the O(N) cost, we assume that scanning for the convergence point adds a small O(\(N^2)\) cost.
Neither of stores is CPU or memory bound with the CPU usage under 25\% and the memory usage under 2\%.

\begin{table}[h!]
\centering
\caption{Performance summary based on Energy and Latency (1-best, 5-worst)}
\begin{tabular}{|l|l|l|l|}
\hline
  Store Type/Email Action & Append & Fetch & Synchronize \\
\hline
  Dovecot & 2 & 4 & 3 \\
  Git & 4 & 3 & 2 \\
  GitGc & 5 & - & 1 \\
  Git-rsync & - & - & 5 \\
  NVCS & 1 & 1 & - \\
  Sqlite & 3 & 2 & 4 \\
\hline
\end{tabular}
\label{t:performance}
\end{table}

\subsection{Energy cost evaluation}
\label{s:energy-eval}

Driven by the social media and video streaming demand, the on-line universe carries enormous amount of data that needs to be stored somewhere, i.e. ubiquitous data centers. Cisco forecast~\cite{Cisco2016} estimates that by 2020 global data center's IP traffic will reach 15.3 zettabytes (ZB) and the data stored will reach 915 exabytes (EB). It is interesting that the amount of data stored on devices will be five times higher than the amount stored in data centers at 5.3ZB. The amount of energy consumed by data centers will triple in the next decade. It presently constitutes 3\% of the global energy usage and contributes 2\% to the green house gas emissions, Bawden~\cite{Bawden2016}. To put things in prospective, 416.2 terawatt hours of electricity that the data centers used in 2015 is significantly higher than the UK total consumption of 300 terawatt hours, and the carbon footprint is on the same level as the airline industry. To control the carbon footprint, companies increase their use of the renewable energy, for instance in 2016 Microsoft announced largest wind purchase to date~\cite{MSWind}. Google used machine learning AI to cut its data centers energy use by 15\%, Vaughan~\cite{Vaughan2016}. To reduce the cooling cost, which contribute 40\% to the overall energy cost, Facebook opened some data centers in Sweden, 70 miles from Arctic Circle, Bawden~\cite{Bawden2016}. Considering that data centers have significant impact on the energy demand, it is reasonable to ask a question of how does the proposed decentralized architecture compare to the centralized data center in terms of the energy cost.

Companies owning data centers generally do not release their detailed energy costs. But as part of the green initiative, in 2011 Google published a white paper outlining the email energy cost comparison of the cloud data center versus local server~\cite{GoogleGreen2011}. Google estimates that their per user email annual energy usage is less than $2.2kWh$.

Radicati email statistics report \cite{Radicati2015} estimates that by 2019 the average number of emails send and received per user per day is 246.5. To estimate the energy cost of the proposed decentralized architecture, I consider the worst case scenario from the energy consumption point of view of GitGc back-end implementation with compression and datasync running on Raspberry Pi 3. The average energy consumed by appending 250 email messages to the back-end is 6,631$\mu Ah$ (Figure \ref{f:appendenergy}). Fetching 250 email messages consumes in average 458$\mu Ah$ (Figure \ref{f:fetchenergy}). GitGc synchronization of 100 messages consumes in average 234$\mu Ah$ (Figure \ref{f:syncenergy}). I estimate the energy consumption for synchronizing 250 messages as $234 * 2.5 = 585\mu Ah$. The total average energy consumption is $6,631 + 458 + 585 = 7,674\mu Ah$. I use Power Usage Effectiveness (PUE)\footnote{\url{https://en.wikipedia.org/wiki/Power_usage_effectiveness}} of 2.5 (highest estimate) in my analysis. I assume that there is a redundant backup device which consumes the same amount of energy. Then the total consumed energy per user per day including the PUE coefficient and the backup is $7,674 * 2.5 * 2 = 38,370\mu Ah$. The nominal Raspberry Pi voltage is $5.1V$. Therefore the consumed energy per day in $Wh$ is $38,370\mu Ah * 5.1V = 195,687\mu Wh$ or $0.195687Wh$. The annual per user energy consumption is $0.195687 * 365 = 71.43Wh$, which is substantially lower than Google's estimate of $2.2kWh$. Note however, that the annual Raspberry Pi 3 energy consumption, depending on the load, is $10(idle)-31(100\%  load)kWh$\footnote{\url{https://raspberrypi.stackexchange.com/questions/5033/}}. Consequently, if a IoT device is under-utilized then the energy consumption is substantially higher than the Google's estimate. One more note is that $2.2kWh$ Google's estimate is 6 years old. Shehabi et al. in \cite[pES-2]{Shehabi2016} show US Data Center total electricity use at 2010 energy efficiency level and current various strategy efficiency level estimates. At 2010 level, the total energy usage in 2017 is about 150 billion $kWh$, and with the best practices it is 40 billion $kWh$. Even if we roughly estimate 4 fold improvement in efficiency then the Google's estimate is $550Wh$, which is still 8 times worse than the projected decentralized email energy usage assuming it is reasonably utilized.

\subsection{Evaluation Summary}
\label{s:summary}

Table ~\ref{t:performance} presents stores performance summary based on the energy and latency evaluation. The evaluation shows that a VCS-like structured store can resource-wise perform at least as good as the base Dovecot IMAP server. However, a VCS implementation can be disk-space and CPU bound when the email messages are committed to disk. Git's snapshot implementation of the revision maintenance has O(\(N^2\)) disk space, energy, and latency cost. Git's Garbage Collection can substantially reduce the disk space overhead at the expense of more energy, latency, CPU, and temporary disk space usage. 

Access to messages is not affected by the type of the store. 

Evaluation of the synchronization protocol shows GitGc has the best performance resource-wise over Dovecot and rsync. This can be explained by the efficient Merkle hash tree detection of the divergence point in the indexed pack file. 
The evaluation further shows that the network IO, disk IO, and compression energy usage, and the latency have a linear bandwidth dependency. The bandwidth can be reduced by compressing the data, transmitting the binary data instead of the base64-encoded data, and de-duplicating attachments. IMAP extensions RFC 4978 and RFC 3516 support compression and binary data transmission. IMAP core RFC 3501 FETCH command supports BODYSTRUCTURE extension data (optional) with the message's body MD5 but doesn't have the option of getting the attachment's MD5. An IMAP extension can be defined to support MIME parts identification via a hash like MD5 or SHA1. This will enable clients supporting this extension to choose whether the attachment needs to be downloaded or not. But IMAP protocol already has rather complicated heuristic based on the mailbox's statistics and metadata to synchronize the client's cache to the server's database. Moreover, supporting version control in IMAP can further complicate the already extension-crowded protocol.

\section{Conclusion}
\label{s:conclusion}

We presented a high level decentralized email architecture that takes advantage of IoT smart-home environment running low-end computing devices with publicly accessible address. The architecture maintains full email history of changes, which is important in present day computing environment where multiple devices, like smartphone, tablet, etc. are accessing and making changes to the email archive and users are unintentionally deleting or misfiling an important email due to the email overload. We also presented a detailed evaluation of latency, CPU, bandwidth, energy, memory, and disk usage for various email stores with the revision control. Our evaluation shows that low-end devices like Raspberry Pi are capable of supporting email architecture with revision-controlled archives and that this architecture can perform at least as good as conventional IMAP server. The architecture proposes to replace legacy IMAP and SMTP protocols with a synchronization protocol, which relies on Merkle hash tree to identify divergence between the local and remote archive.
	
We see emerging IoT technology as a promising platform not just for the email but for the overall unified messaging architecture where the data is put back under user's control. We only touched just a few aspects of this architecture but we hope it can encourage more research in this area. Future research can focus on an optimal structure to maintain email revisions. Details of the synchronization protocol have to be looked at, for instance how to prioritize MIME parts download so that mobile or bandwidth-limited devices can have the email preview without downloading the whole message. Establishing connection between peers is a big research area with one thought to use existing email providers as the DNS server. And last but not least, user's privacy in IoT environment is a substantial research subject.

\bibliographystyle{ACM-Reference-Format}
\bibliography{paper}
\end{document}